\documentclass[a4paper,11pt]{article}
\usepackage{pos}
\usepackage{threeparttable}
\usepackage{hyperref}
\usepackage{amssymb}
\usepackage{adjustbox}

\title{Probing Neutrino Emission from X-ray Blazar Flares observed with Swift-XRT}
 \ShortTitle{Neutrino Emission from X-ray Flares}

\author*[1]{Stamatios~I. Stathopoulos}
\author[1]{Maria~Petropoulou}
\author[2,3,4]{Paolo~Giommi}
\author[5]{Georgios~Vasilopoulos}
\author[6,7]{Paolo~Padovani}
\author[1]{Apostolos~Mastichiadis}

\affiliation[1]{Department of Physics, National and Kapodistrian University of Athens,\\ University Campus Zografos, GR 15783, Greece}
\affiliation[2]{
Institute for Advanced Study, Technische Universität München,\\ Lichtenbergstrasse 2a, D-85748 Garching bei München, Germany}
\affiliation[3]{Associated to Agenzia Spaziale Italiana, ASI, via del Politecnico s.n.c.,\\ I-00133 Roma Italy}
\affiliation[4]{ICRANet, Piazzale della Repubblica 10,\\ I-65122, Pescara, Italy}
\affiliation[5]{Université de Strasbourg, CNRS, Observatoire astronomique de Strasbourg,\\ UMR 7550, 67000, Strasbourg, France}
\affiliation[6]{European Southern Observatory, Karl-Schwarzschild-Str. 2,\\ D-85748 Garching bei München, Germany}
\affiliation[7]{Institute for Advanced Study, Technische Universität München, Lichtenbergstrasse 2a,\\ D-85748 Garching bei München, Germany}

\emailAdd{stamstath@yahoo.gr}

\abstract{Blazars are the most extreme subclass of active galactic nuclei with relativistic jets emerging from a super-massive black hole and forming a small angle with respect to our line of sight. Blazars are also known to be related to flaring activity as they exhibit large flux variations over a wide range of frequency and on multiple timescales, ranging from a few minutes to several months. The detection of a high-energy neutrino from the flaring blazar TXS~0506+056 and the subsequent discovery of a neutrino excess from the same direction have naturally strengthened the hypothesis that blazars are cosmic neutrino sources. While neutrino production during gamma-ray flares has been widely discussed, the neutrino yield of X-ray flares has received less attention. Motivated by a theoretical scenario where high energy neutrinos are produced by energetic protons interacting with their own X-ray synchrotron radiation, we make neutrino predictions over a sample of a sample of X-ray blazars. This sample consists of all blazars observed with the X-ray Telescope (XRT) on board Swift more than 50 times from November 2004 to November 2020. The statistical identification of a flaring state is done using the Bayesian Block algorithm to the 1 keV XRT light curves of frequently observed blazars. We categorize flaring states into classes based on their variation from the time-average value of the data points. During each flaring state, we compute the expected muon plus anti-muon neutrino events as well as the total signal for each source using the point-source effective area of Icecube for different operational seasons. We find that the median of the total neutrino number (in logarithm) from flares with duration $<30$~d  is  $\mathcal{N}^{(\rm tot)}_{\nu_{\mu}+\bar{\nu}_{\mu}} \sim 0.02$.
}

\DeclareRobustCommand{\VAN}[3]{#2}
\let\VANthebibliography\thebibliography
\def\thebibliography{\DeclareRobustCommand{\VAN}[3]{##3}\VANthebibliography}

\FullConference{37$^{\rm{th}}$ International Cosmic Ray Conference (ICRC 2021)\\
		July 12th -- 23rd, 2021\\
		Online -- Berlin, Germany}

\begin{document}
\maketitle

\section{Introduction}
Blazars are the most extreme subclass of active galactic nuclei with relativistic jets emerging from a super-massive black hole and forming a small angle with respect to our line of sight \citep{Urry_1995}. The spectral energy distribution of these objects exhibits a double-hump shape ranging from radio wavelengths up to high-energy $\gamma$-rays \citep{Padovani_2017}.  
\par   In 2017 the IceCube Neutrino Observatory detected a high energy neutrino with deposited energy at the detector of $\sim 24$ TeV while the reconstruction of the event indicates that the most probable neutrino energy was 290 TeV \citep{2018Sci...361.1378I}. Since neutrinos have a neutral charge we can determine the location of production. The arrival of the detected high-energy neutrino was in spatial coincidence with a known blazar TXS 0506+056, which was at a 6-month $\gamma$-ray flaring state at this moment, and provided the first 3$\sigma$ neutrino source \citep{2018Sci...361.1378I}. The same blazar seems to also be related to a high-energy neutrino excess with respect to the atmospheric background during the 2014/15 period  \citep{2018Sci...361.1378I}.  During the latter period, TXS 0506+056  was not undergoing any significant variability or flaring in any observed wavelength.  
\par 

There have been many theoretical attempts trying to connect high-energy neutrino production with $\gamma$-ray flares from blazars \citep{2019ICRC...36.1038Y,2020ApJ...893..162F} . In all cases, an accelerated hadronic population inside the jet is needed in order to produce high-energy neutrinos via the decay of charged pions produced in  photomeson interactions. In this work, we present quantitative neutrino predictions of the hadronic X-ray flaring scenario of blazars. We compute the number of muon and antimuon neutrinos above 100~TeV expected for IceCube from X-ray flares of blazars that were observed  more than 50 times with the X-ray Telescope  (XRT, \citep{xrt2005}) on board the Neil Gehrels \emph{Swift} Observatory between November 2004 and November 2020 (Giommi et al., 2021, submitted).

\section{Theoretical Model}\label{sec:Theory}
The basic assumption in our model is that every X-ray flare is the product of an accelerated population inside the blazar's jet. We assume that these relativistic protons lose energy via synchrotron radiation and photomeson interactions. The target photon field is the synchrotron radiation emitted by the relativistic protons. Protons with Lorentz factor that exceed the energy threshold for pion production will produce neutral and charged pions. The latter will decay into lighter leptons, including muon and electron neutrinos (and antineutrinos). The characteristic neutrino energy generated from these protons will be approximately $5\%$ of the parent proton energy.
\begin{eqnarray}
\label{eq:Enu}
\varepsilon_{\nu} &\simeq& 0.05 \,  \mathcal{D}(1+z)^{-1}\gamma'_{\rm p}m_{\rm p} c^2 \\ \nonumber 
& \simeq & 0.6 \, \sqrt{\mathcal{D}_1 B_1^{'-1}\varepsilon_{\rm keV}(1+z)^{-1}}~{\rm PeV}
\end{eqnarray}
where $\gamma'_{\rm p}$ is proton Lorentz factor, $B'$ is the magnetic field and $\varepsilon_{\rm keV}$ is the characteristic observed synchrotron energy normalized at 1~keV. Primed quantities are measured in the rest frame of the emission region, while unprimed quantities correspond to the measurements in the observer's frame.
\par  Here, we focus on the “neutrino-rich” scenario where the photomeson energy loss rate is comparable to the energy loss rate due to proton-synchrotron radiation $t'^{-1}_{\rm syn}\simeq t'^{-1}_{\rm mes}$ \citep[for details see][]{2021ApJ...906..131M}. During the peak of the flaring state, the X-ray luminosity is closely related to the neutrino luminosity with a scaling factor, i.e. $L_{\rm X}=\xi_{\rm X} L_{\nu +\bar{\nu}}$, where $\xi_{\rm X}\approx1$. 
\par The differential neutrino plus antineutrino energy flux of all flavours can be modeled as a power law (with power law index $s(t)$) with an exponential cutoff at the characteristic neutrino energy $\varepsilon_{\nu, \rm c}$\footnote{To determine the characteristic neutrino energy we substitute $\varepsilon_{\rm keV}$ with the peak energy of the X-ray spectrum in $\varepsilon F_{\varepsilon }$}.
\begin{equation}
F_{\nu+\bar{\nu}}(\varepsilon_{\nu},t) = F_0(t)\varepsilon_{\nu}^{-s(t)} e^{-\varepsilon_{\nu}/\varepsilon_{\nu, \rm c}}
\label{eq:Fnu}
\end{equation}
where $F_0(t)$ is the time-dependent normalization factor which is determined using $\int_{\varepsilon_{\nu,\min}}^{\varepsilon_{\nu,\max}} {\rm d}\varepsilon_{\nu}F_{\nu+\bar{\nu}}(\varepsilon_{\nu},t)=\xi_{\rm X} \int_{\varepsilon_{\min}}^{\varepsilon_{\max}} {\rm d}\varepsilon \, F_{\rm X}(\varepsilon,t)$.
\par
Therefore, the expected number of muon plus antimuon neutrinos from an X-ray flare can be calculated as
\begin{equation}
\mathcal{N}_{\nu_{\mu}+\bar{\nu}_{\mu}}=\frac{1}{3}\int_{t_{\rm ini}}^{t_{\rm end}}{\rm d}t \int_{E_{\nu, \min}}^{E_{\nu, \max}}{\rm d}\varepsilon_{\nu} \, A_{\rm eff}(\varepsilon_{\nu},\delta) \frac{F_{\nu+\bar{\nu}}(\varepsilon_{\nu},t)}{\varepsilon_{\nu}},
\label{eq:events}
\end{equation}
where we assumed vacuum neutrino mixing and used 1/3 to convert the all-flavour to muon neutrino flux. Moreover, $t_{\rm ini}$ and $t_{\rm end}$ define the duration of the X-ray flare as $\Delta t = t_{\rm end}-t_{\rm ini}$ and $A_{\rm eff}(\varepsilon_{\nu},\delta)$ is the energy-dependent and declination-dependent point-source effective area of IceCube \citep[]{Aartsen_2020, 2021arXiv210109836I} with respect to each generation of Icecube. For the integration over energies we set $E_{\nu, \min}=100$~TeV in order to exclude the contribution of the atmospheric background. As for the maximum energy, we use the maximum energy to which $A_{\rm eff}$ is computed as $E_{\nu, \max}$.  
 \begin{figure}
    \centering
    \includegraphics[width=0.55\textwidth]{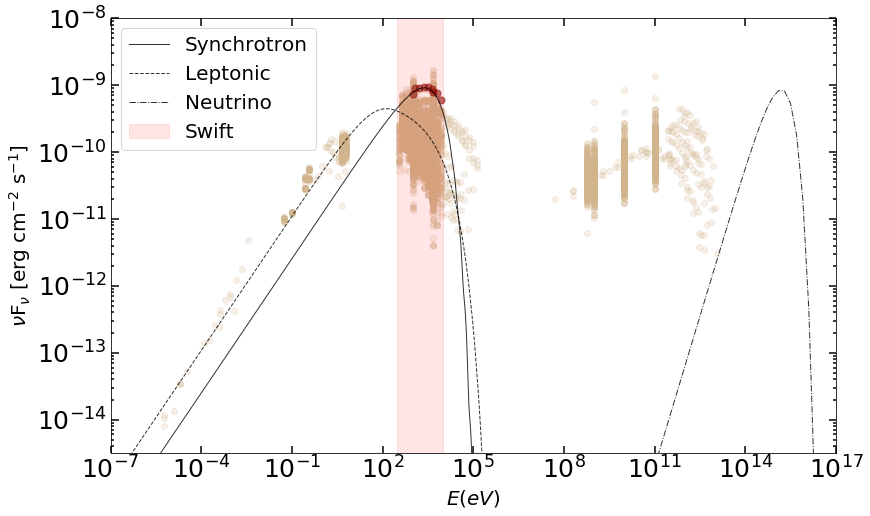}
    \caption{Spectral energy distribution of Mkn~421 compiled using data from various instruments and epochs (adopted from the \href{http://openuniverse.asi.it/ouspectra/catalog_description.html}{Open Universe for Blazars}). The spectrum of an X-ray flare is highlighted with red symbols and the shaded region indicates the 0.5-10 keV energy range. Solid and dash-dotted lines present the proton synchrotron  spectrum and the accompanying all-flavour neutrino spectrum of the flare, respectively. A likely contribution to the non-flaring spectrum from an accelerated electron population is also displayed (dotted line).}
    \label{fig:model_schem}
\end{figure}
\section{X-ray light curves and flaring states}\label{sec:l-c_f-s}

\begin{figure*}[ht]
\includegraphics[width=0.45\textwidth]{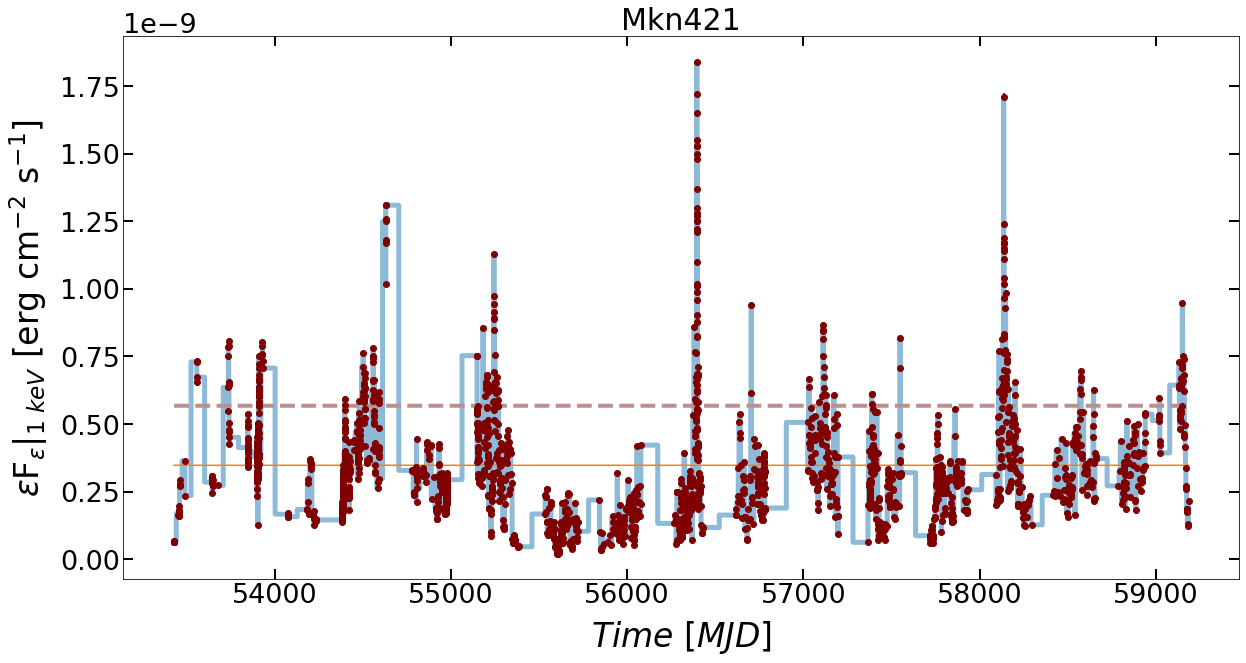}
\hfill
\includegraphics[width=0.42\textwidth]{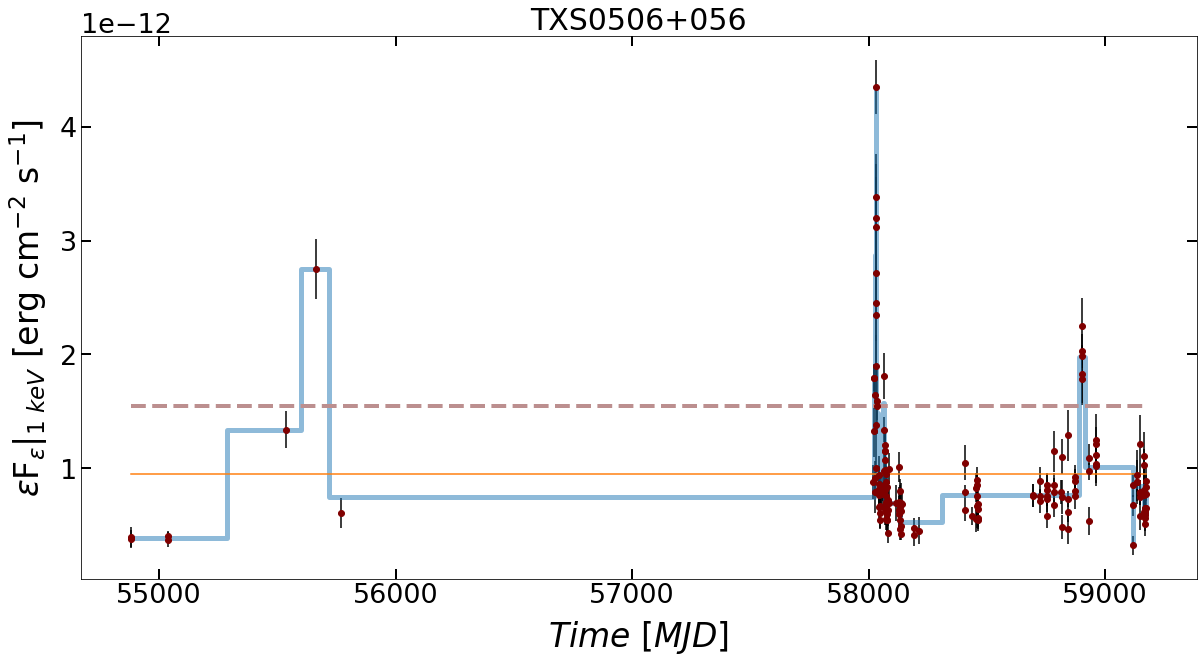}
\caption{1~keV light curves of two blazars from our sample (symbols): Mkn 421 and TXS~0506+056 (neutrino candidate). Error bars indicate the 68\% uncertainty in flux. Solid lines show the Bayesian block representation of the light curves. Long horizontal lines with no sampling between a data point and a new block do not guarantee a stable flux. The orange solid line indicates the mean value $\mu$ of all flux measurements while the dashed line correspond to $\mu+\sigma$. }
\label{fig:1keV}
\end{figure*}

For our analysis, we used data from the X-ray telescope (XRT, \cite{xrt2005}) onboard the Neil Gehrels Swift Observatory obtained during the period November 2004 to November 2020 (14 years). More specifically, we used all blazars that have been observed at least 50 times in this period with Swift. To look for variability we used the 1 keV X-ray light curves as obtained by Giommi et al., 2021, submitted. 1~keV fluxes are calculated from the best-fit power-law model to the data or from converting the broadband count rate to energy flux for observations with less than 20 counts. Figure~\ref{fig:1keV} shows two indicative light curves of the sample.

\par  Because of the irregular sampling of the data, we applied the Bayesian block algorithm described in \citep{Scargle:2013} to the 1 keV XRT light curves of frequently observed blazars to characterize statistically significant variations, at the same time suppressing the inevitable contaminating observational errors. Because of the nature of data ("point measures"), the only free parameter of the algorithm is $p_0$ which gives the false alarm probability to compute the prior on the number of bins. We set $p_0=0.1$ throughout all calculations. In Fig.~\ref{fig:1keV} the Bayesian block representation of the light curves is overplotted with solid lines while the height of each block is the weighted average of all flux measurements belonging to it. 
\par
The flaring states are defined based on how many $n\sigma$ above the mean value ($\mu$) of all flux measurements the flux block lies. Number $n$ could be any real number depending on the variability of each source and $\sigma$ is the standard deviation of the flux measurements. Besides the flux of a block, the Bayesian blocks algorithm returns the duration of the block $\Delta t$,  which we consider to be the duration of the X-ray flare used in calculations of the expected neutrino events. 
We classify flares in two types based on the block flux, $f_{\rm B}$, as follows
\begin{itemize}
\centering
    \item Type A: $\mu+\sigma < f_{\rm B} < \mu+3\sigma$  
    \item Type B: $f_{\rm B} > \mu+3\sigma $  
\end{itemize}

\par
 By performing a two-sample Kolmogorov-Smirnov test on the distributions of $\log(\Delta t)$ and $\log(f_{\rm B})$ of flares belonging in different types we can reject the hypothesis that these originate from the same population. 
More specifically, we find that Type B flares are brighter and shorter in duration than Type A flares. 
Some large gaps inside the light curves may create false flaring states. After visual inspection of the light curves we find that flaring states with $\Delta t>60$ days contains $\sim 1-2$ XRT flux measurements. During these long-duration blocks the exact behavior of the light curve cannot be predicted. Using the flux of a couple XRT snapshots with total duration of a few ks as a proxy for the source flux state on week-long or even month-long periods introduces big uncertainties in the predicted neutrino fluence. Hence, if the block duration is $\Delta t>10$ d and contains only one XRT observation, we set $\Delta t=1$ d, which is close to the most probable value of the duration distribution of the flaring states.

\section{Neutrino predictions from the X-ray flares}\label{sec:predictions}

\begin{figure*}[ht]
\centering
\includegraphics[width=0.6\textwidth]{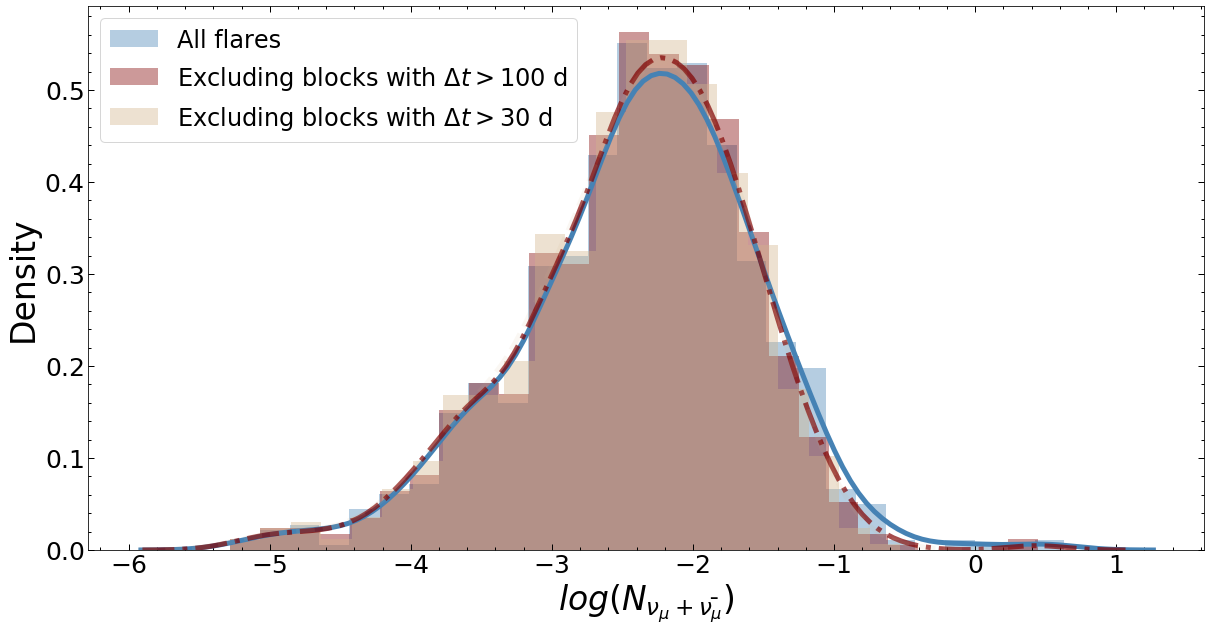}
\caption{Normalized distribution of the expected muon and anti-muon neutrino number from X-ray flares (in logarithmic scale). Distributions obtained after removing blocks with $\Delta t > 100$~d and 30~d are overplotted for comparison.}
\label{fig:1Dhisto-neutrinos}
\end{figure*}

Figure \ref{fig:1Dhisto-neutrinos} shows the distribution of the predicted number of muon and antimuon neutrino events from all X-ray flares (blue line). The median of the distribution is close to $\sim 0.02$ events. The same figure contains the histograms of $\log( \mathcal{N}_{\nu_{\mu}+\bar{\nu}_{\mu}})$ after excluding the contributions of flaring states (blocks) with $\Delta t > 100$~d (maroon) and 30~d (tan). The distributions of the $\log( \mathcal{N}_{\nu_{\mu}+\bar{\nu}_{\mu}})$ using these cuts in duration does not change their shape neither their mean and median values. 
Therefore the contribution of long-duration blocks to the neutrino events of our sample is small. In fact, the majority of the predicted neutrino events in the sample originates from flares with $\Delta t\sim 1-10$~d.
\par  In our model the duration of the neutrino flare is similar to the duration of the X-ray flare, since the relevant proton cooling timescales are comparable.  Hence, flares with a longer duration produce a larger number of muon and antimuon neutrino events. 
For a given flare duration, flares with higher X-ray fluxes are also found to produce a higher number of events. 
\par
\begin{figure*} 
\centering
\includegraphics[width=0.75\textwidth]{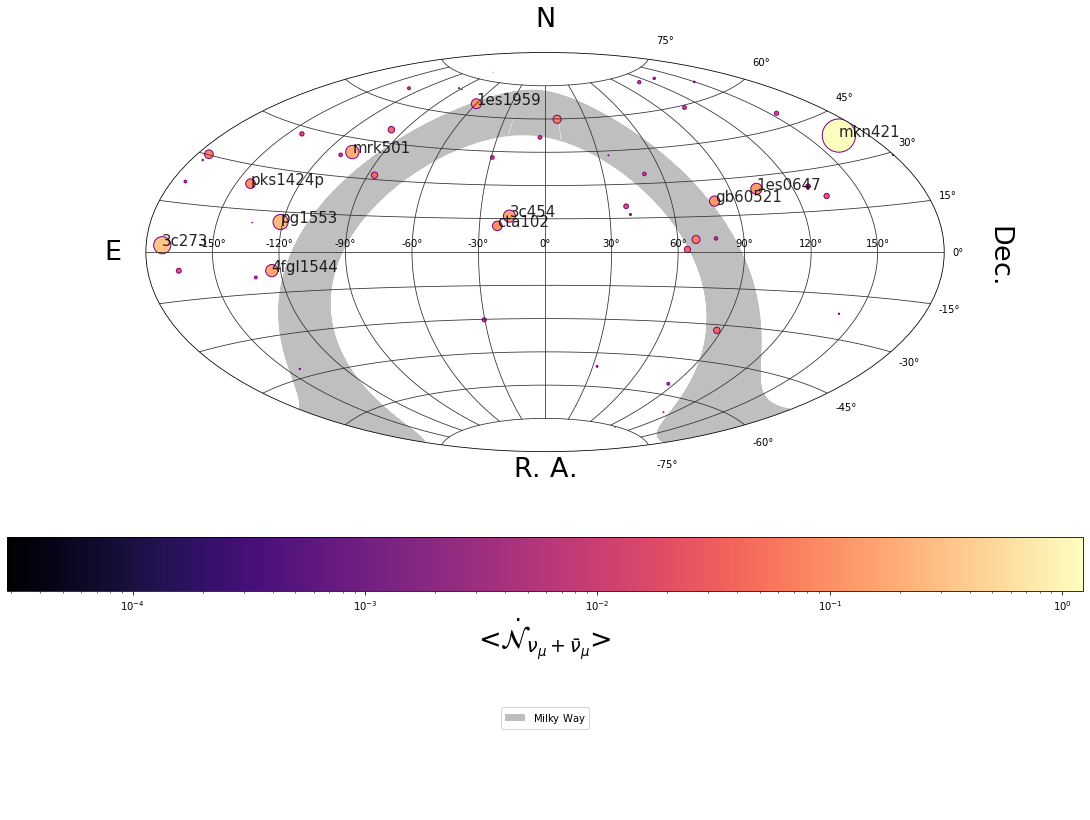}
\caption{All-sky plot in equatorial coordinates showing the yearly average neutrino rate of sources due to X-ray flares. The listed Blazars have a yearly average neutrino rate greater than 0.1 yr$^{-1}$ from flares with $\Delta t <30$~d. The grey region represents the Milky Way plane.}
\label{fig:eq_co}
\end{figure*}

By adding up all muon and antimuon neutrinos predictions from all individual flares we can find the yearly average neutrino rate of each source. Figure \ref{fig:eq_co} shows a sky map with the locations of all blazars from our sample indicated with circles. The symbol size and color corresponds to the yearly average neutrino rate of muon and antimuon neutrinos (see color bar and inset legend). 
The sources displayed on the map have 
$\dot{\mathcal{N}}^{\rm (tot)}_{{\nu}_{\mu}+\bar{\nu}_{\mu}}> 0.1$ yr$^{-1}$ from flares with $\Delta t <30$~d.

\section{Effects of model parameters and source declination}
Up to this point we have presented results for fixed values of the magnetic field strength ($B'=10$~G) and Doppler factor ($\mathcal{D}=10$) in all sources. In reality, the magnetic field is going to affect the predictions since higher values of the magnetic field strength $B'$ would lower the proton Lorentz factor $\gamma'_{\rm p}$ needed to produce synchrotron photons of energy $\varepsilon_{\rm pk}$. For sufficiently strong magnetic fields, it is therefore possible that the proton Lorentz factor drops below the threshold value for pion production on synchrotron photons of the same energy. Replacing all variables that are related to the magnetic field in equation~(\ref{eq:events}) we get
\begin{equation}
	\mathcal{N}_{\nu_{\mu}+\bar{\nu}_{\mu}}(B',\delta)\propto B^{'(1-s)/2}\int_{E_{\nu,\min}}^{E_{\nu,\max}}{\rm d}\varepsilon_{\nu} \varepsilon_{\nu}^{-s-1} \, A_{\rm eff}(\varepsilon_{\nu},\delta) e^{-a\varepsilon_{\nu}B^{'1/2}},
\label{eq:events3}
\end{equation}
where $\delta$ is the source declination angle, $a$ is parameter depending on the Doppler factor and source redshift. 
\par
\begin{figure*}[ht]
\centering
\includegraphics[width=0.7\textwidth]{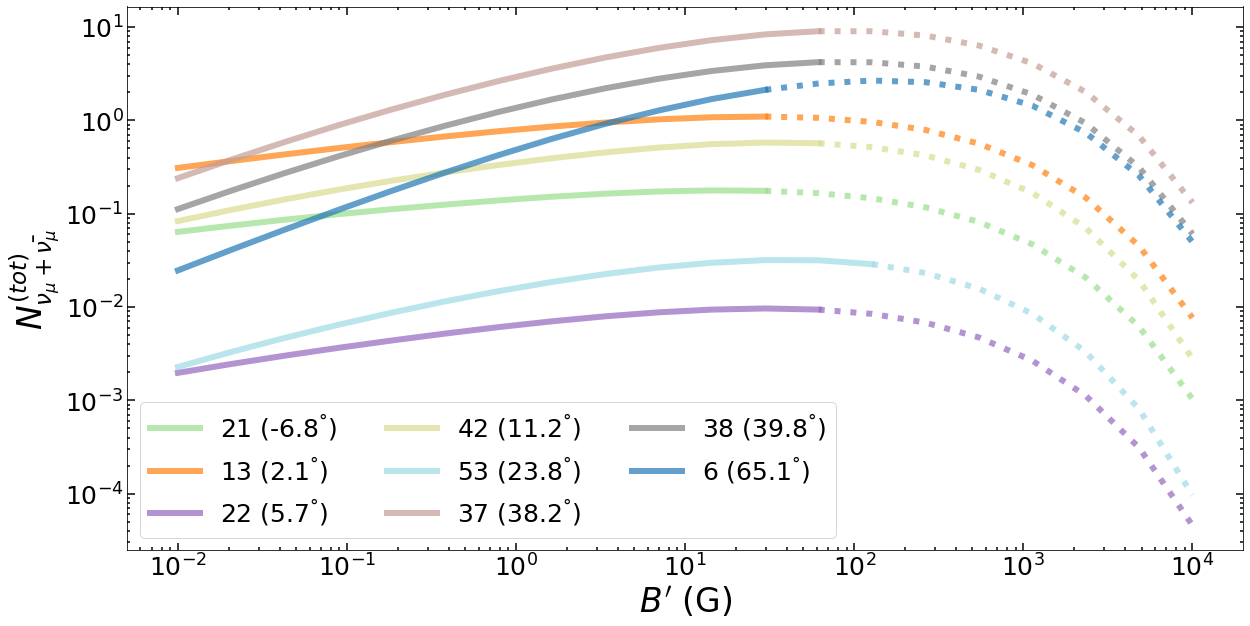}
\caption{Stacked number of muon and antimuon neutrinos from flares with $\Delta t<30$ d as a function of the magnetic field strength fora few indicative sources (see inset legend; values in the parenthesis indicate declination angles). Other parameters used here are: $\mathcal{D}=10$ and $\varepsilon_{\rm pk=}1keV $}
\label{fig:Stacked_vs_B}
\end{figure*}

 
 To derive an analytical expression for the predicted muon plus antimuon neutrino number versus the magnetic field, an analytical expression for the effective area for a given declination is necessary. Still, eq.~\ref{eq:events3} can be used to understand the dependence on $B'$ and $\delta$ in two limiting regimes. 
More specifically, for small values of the magnetic field ($B'\ll10(0.6~{\rm PeV}/\varepsilon_{\nu, \rm pk})^2\mathcal{D}_1\varepsilon_{ \rm keV}(1+z)^{-1}~{\rm G}$, the exponential in eq.~\ref{eq:events3} can be neglected and $\mathcal{N}_{\nu_{\mu}+\bar{\nu}_{\mu}} \propto B^{'(1-s)/2}$. For higher values of $B'$, the exponential dependence will dominate. Consequently, there is a critical value of the magnetic field for each source that maximizes the predicted neutrino number. This critical value depends on the source declination through $A_{\rm eff}$. Figure \ref{fig:Stacked_vs_B} shows this dependence of the total neutrino number (after excluding long-duration blocks with $\Delta t >30$~d) for a few indicative sources (see also Table\ref{table_1}). Solid lines indicate magnetic field values for which the proton Lorentz factor exceeds the energy threshold for pion production on X-ray photons, while dotted lines are used otherwise. 

\par
The Doppler factor of each source $\mathcal{D}$ is also a parameter which will affect the results. The predicted neutrino number has the same dependence on the Doppler factor $\mathcal{D}$ as well as with the strength of the magnetic field $B'$. Its effects, however, are less pronounced in the range of values expected for blazar jets compared to the strength of the magnetic field.

\begin{table*}
\centering
\caption{Sample of blazars observed more than 50 times with Swift/XRT and model predictions about total number and average yearly rate of muon and antimuon neutrinos expected to be detected by IceCube.}
\begin{adjustbox}{width=1\textwidth}

\begin{threeparttable}
\begin{tabular}{*{8}{c}}
\hline 
Source index & Source name & Dec (deg) & Class  & $\mathcal{N}^{(\rm tot)}_{\nu_{\mu}+\bar{\nu}_{\mu}}$ & $\mathcal{N}^{(\rm tot)}_{\nu_{\mu}+\bar{\nu}_{\mu}}(\Delta t<30~{\rm d})$ &  $\langle \dot{\mathcal{N}}_{\nu_{\mu}+\bar{\nu}_{\mu}} \rangle$ ($\times10^{-4}~\rm yr^{-1}$) & $ \dot{\mathcal{N}}^{\rm (atm)}_{\nu_{\mu}+\bar{\nu}_{\mu}}$ ($\times10^{-4}~\rm yr^{-1}$) \\ 
\hline 
6 & 1ES~1959+650 & 65.15  &	 HSP  &$0.81 \pm  0.05$ &  $0.81  \pm  0.05 $ & $ 1101.4  \pm  68.4    $&$ 4.0$\\ 
13 & 3C~273& 2.05&     LSP  &$1.7 \pm  0.1$ &  $0.80 \pm  0.07$ & $ 3286.1  \pm  296.4   $&$ 15.9$\\
21 & 4FGL~J1544.3-0649 & -6.82 &  HSP  &$0.15  \pm  0.03$ &  $0.15 \pm  0.03$ & $ 1665.5  \pm  328.6   $&$ 9.0$\\  
22 & TXS~0506+056 & 5.69 &   ISP &$0.013  \pm  0.004$ &  $0.013 \pm  0.004$ & $ 127.5   \pm  38.8    $&$ 14.8$\\    
37& Mkn~421& 38.21 & HSP &$9.6  \pm  0.2$ &  $4.2 \pm  0.1$ & $ 12284.7 \pm  345.5   $&$ 8.4$\\
38& Mkn~501& 39.76 &  HSP &$1.38 \pm  0.05$ &  $1.37 \pm  0.05$ & $ 1986.8  \pm  79.3    $&$  8.4$\\ 
42& PG~1553+113 & 11.19 &   HSP  &$0.77  \pm  0.07$ &  $0.57 \pm  0.05$ & $ 2543.8  \pm  204.1   $&$ 13.9$\\      
53& PKS~1424+240 & 23.80 &  ISP &$0.18 \pm  0.03$ &  $0.031 \pm  0.003$ & $ 1033.1  \pm  94.6    $&$ 11.0$\\\hline 
\end{tabular}
\end{threeparttable}
\label{table_1}
\end{adjustbox}
\end{table*}
\bibliographystyle{mnras}
\bibliography{bibliography} 
\end{document}